\documentclass[a4paper,11pt]{article}
\usepackage{pos}

\title{The Influence of Satellite Trails on H.E.S.S. Gamma-Ray Astronomical Observations}
 \ShortTitle{Satellite Trails in H.E.S.S. Data}

\author*[a,b]{Samuel T. Spencer}
\author[a]{Thomas Lang}
\author[a]{Alison M.W. Mitchell}

\affiliation[a]{Friedrich-Alexander-Universit{\"a}t Erlangen-N{\"u}rnberg, Erlangen Centre for Astroparticle Physics, Nikolaus-Fiebiger-Str. 2, D 91058 Erlangen, Germany}

\affiliation[b]{Department of Physics, Clarendon Laboratory, Parks Road, Oxford, OX1 3PU, United Kingdom}

\emailAdd{samuel.spencer@fau.de}

\abstract{The number of satellites launched into low earth orbit has almost tripled (to over 4000) in the last three years due to the increasing commercialisation of space. Satellite constellations with a total of over 400,000 satellites are proposed to be launched in the near future. Many of these satellites are highly reflective, resulting in a high optical brightness that affects ground-based astronomical observations across the electromagnetic spectrum. Despite this, the potential effect of these satellites on Imaging Atmospheric Cherenkov Telescopes (IACTs) has so far been assumed to be negligible due to their nanosecond integration times. This has, however, never been verified. We aim to identify satellite trails in data taken by the High Energy Stereoscopic System (H.E.S.S.) IACT array in Namibia, using Night Sky Background (NSB) data from the CT5 camera installed in 2019. We determine which observation times and pointing directions are affected the most, and evaluate the impact on Hillas parameters used for classification and reconstruction of high-energy Extensive Air Shower events. Finally, we predict how future planned satellite launches will affect gamma-ray observations with IACTs.}

\ConferenceLogo{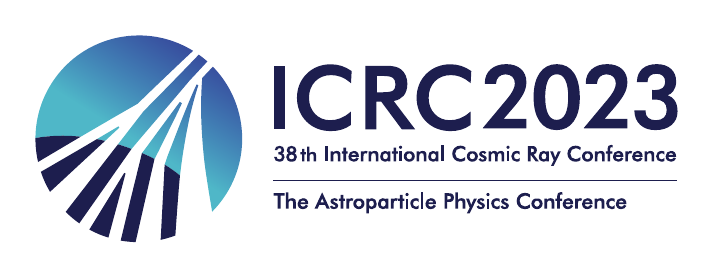}

\FullConference{%
38th International Cosmic Ray Conference (ICRC2023)\\
  26 July - 3 August, 2023\\
  Nagoya, Japan}

\begin{document}
\maketitle
\section{Introduction}

The advent of Starlink and other similar satellite mega-constellations has drastically increased the number of highly reflective satellites in Earth orbit in recent years, with at least four companies planning $\gtrsim10^3-10^4$ total launches each in the near future. This poses a potential threat to the viability of ground-based astronomical observations across the electromagnetic spectrum \cite{Hainaut}. However, despite observing Cherenkov light which peaks in the optical to near-UV frequency range, and being exposed to the night sky, the effect of such satellites on gamma-ray-observing Imaging Atmospheric Cherenkov Telescopes (IACTs) has never been quantified with experimental data. This is potentially due to their short ($\mathrm{\sim10\,ns}$) signal integration times, but photons from satellites could still trigger IACTs or bias the reconstruction of air shower events. As such, should the current trend of increasing launches continue, the next-generation Cherenkov Telescope Array (CTA) observatory could potentially be non-trivially affected in the 2030s. Our work aims to identify the presence of satellite trails in data from the current generation High Energy Stereoscopic System (H.E.S.S.) IACT array, quantify their impact on standard event classification and reconstruction techniques, and determine if additional mitigation strategies are required for the future. These proceedings serve as a brief summary of \cite{Langetal}.

H.E.S.S. is an array of five IACTs located in the Khomas Highland of Namibia that has been in operation since 2004. It consists of four 12\,m diameter IACTs (CT1-4) and a fifth 28\,m diameter instrument (CT5) that was added in 2012. Each H.E.S.S. telescope is equipped with a sensitive, high-speed camera with a focal plane instrumented with photomultipliers. The camera of CT5 has recently been replaced with a version of the FlashCam camera \cite{FlashcamPerformance}, with the aim of improving stability of observations and reducing the energy threshold of the instrument. H.E.S.S. and all other IACTs operate by observing the Cherenkov light emitted due to energetic particles in Extensive Air Showers (EAS) induced by Very-High-Energy (VHE) gamma-rays interacting in the atmosphere. The primary particle type, energy and incident direction are typically inferred by extracting so-called Hillas parameters from IACT images observing this EAS stereoscopically. Multivariate and template-based analysis techniques are then typically applied to the data following this step in order to improve performance (see e.g. \cite{parsons}). 

\section{Method}
\begin{figure}[t]
    \centering
    \includegraphics[width=0.8\hsize]{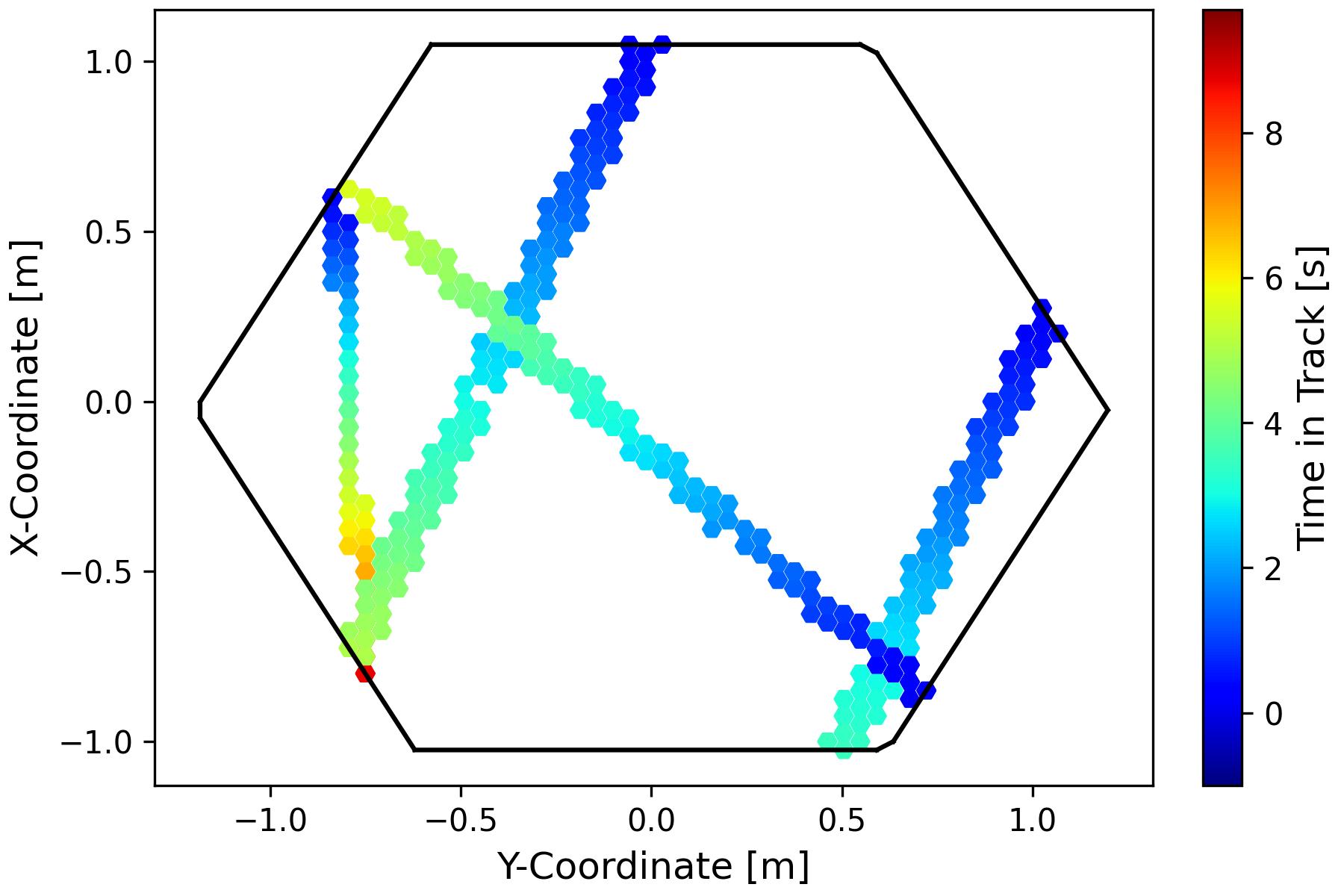}
    \caption{An example of an observing run in which we detect multiple trails in the CT5 camera. }
    \label{fig:Camera_multiple_trails}
\end{figure}

The presence of satellites crossing the field of view of IACTs is easily inferred due to the deactivation of pixels as a safety measure in response to a high incident photon flux. However, this is naturally limited to the very brightest satellites. We have developed a new method of extracting satellite trails from FlashCam measurements of the Night Sky Background (in MHz) across the camera plane (taken concurrently with EAS observations) which appears to be able to detect dimmer trails. For the older CT1-5 cameras these NSB measurements have been typically performed by analysing measurements of the photomultiplier currents. As FlashCam-type cameras are DC-coupled, the NSB rate can be measured using the DC baseline level. Crucially, this can be performed with FlashCam-type cameras in near-real-time (every 0.1\,s). This allows for satellite trails to be identified in this NSB data (unlike the older cameras) as this is much shorter than the typical duration of a satellite trail crossing the H.E.S.S. field of view.

We consider approximately three years of H.E.S.S. observations since the FlashCam CT5 upgrade, and apply standard H.E.S.S. observation selection criteria; partly to ensure instrument stability, and partly because we are concerned with trails that would contaminate typical astronomical observations. We also exclude $5^{\circ}$ regions around the Large Magellanic Cloud and the Colliding Wind Binary $\eta$ Carinae, as they are abnormally bright regions of the sky in the optical and near-UV bands \cite{etacar}.  Trails can then be identified in the FlashCam NSB data by performing a series of data cuts (such as the number of unique NSB map entries above a 900MHz NSB rate in a 28\,min observing run) in combination with a pixel-neighbourhood finder, which appends pixels to trail objects if they pass the cuts. Additional selection is then performed on the candidate trails to eliminate, for example, short-duration trails more likely to be associated with meteors rather than satellites. This procedure results in 1658 trail detections in this dataset. An example of a 28-minute observing run with multiple such candidate trails can be seen in Figure \ref{fig:Camera_multiple_trails}. We observed multiple runs such as this where there are trails in parallel at differing times; this is likely the result of the launch of satellite constellations in `trains' \cite{Hainaut}.

\section{Results}
\begin{figure}
\resizebox{\hsize}{!}
        {\includegraphics{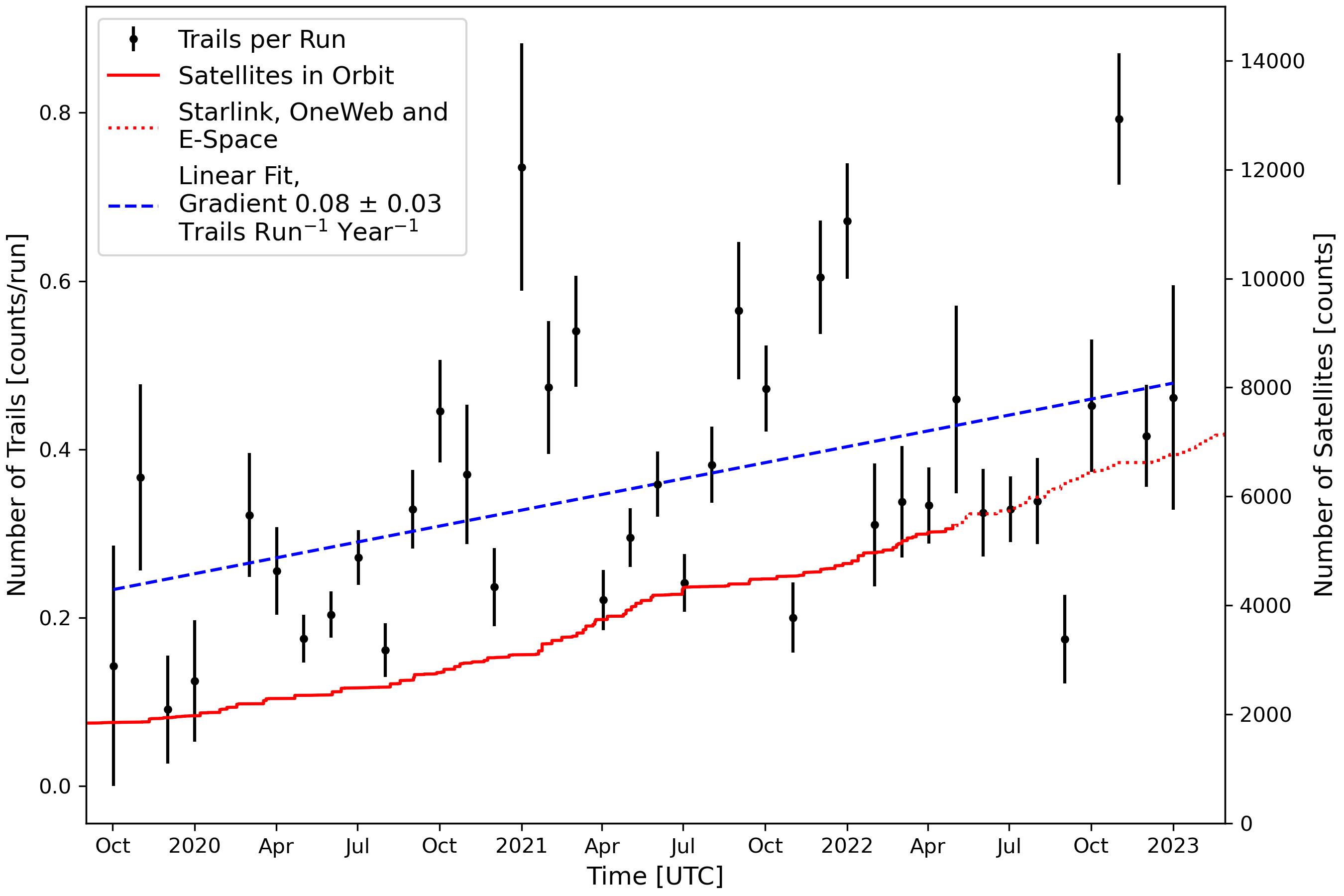}}
    \caption{The number of trails we detect over the three years of data we consider with their associated Poisson error. A linear fit is performed to this data which shows an increasing trend over time. Orbital and satellite launch data is also shown for comparison, with which our trail detections are mildly correlated.}
    
         \label{fig:monthly_avgerge}
\end{figure}

\begin{figure}
\centering
  \centering
  \includegraphics[width=0.48\textwidth]{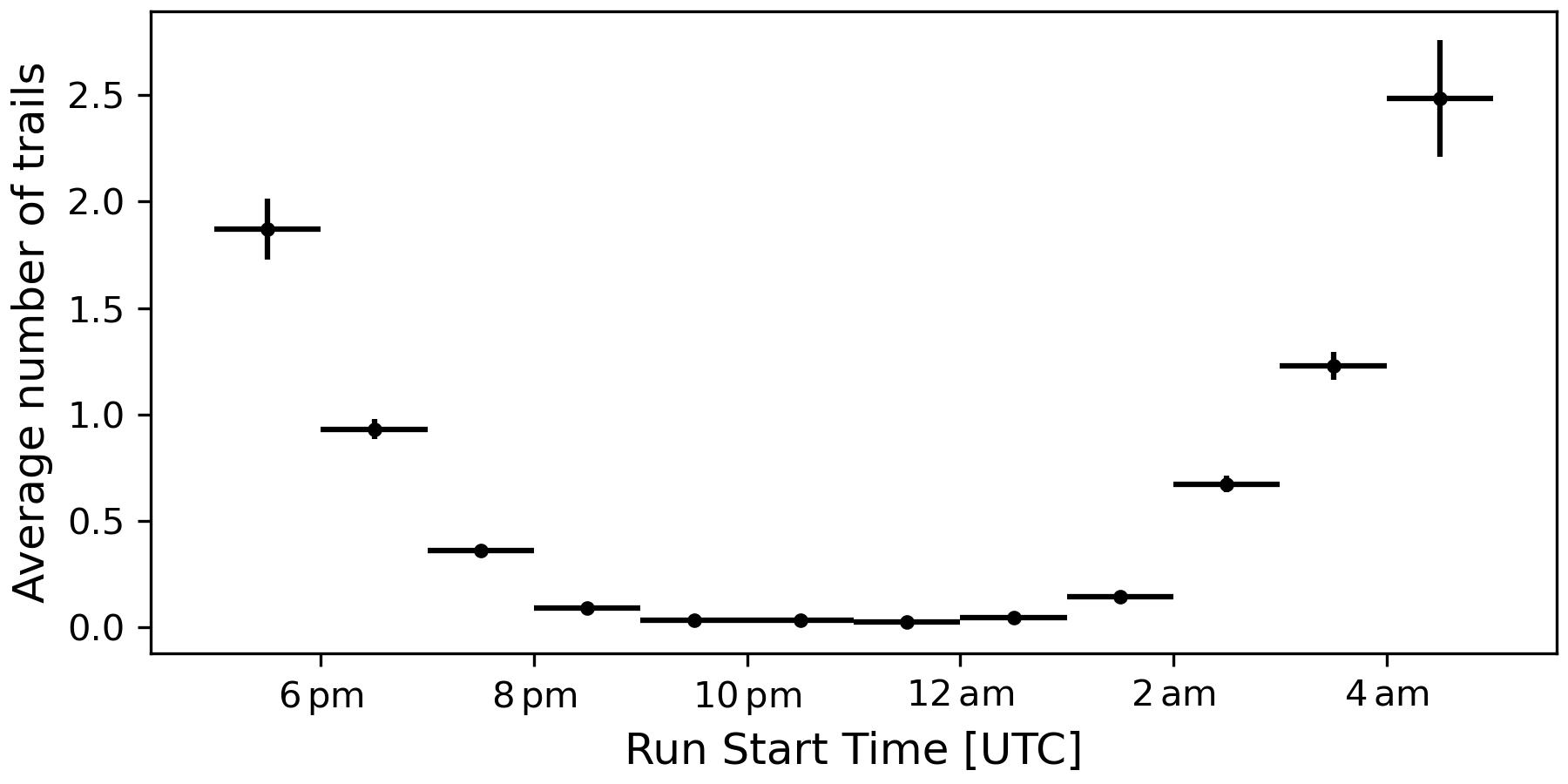}
  \includegraphics[width=0.48\textwidth]{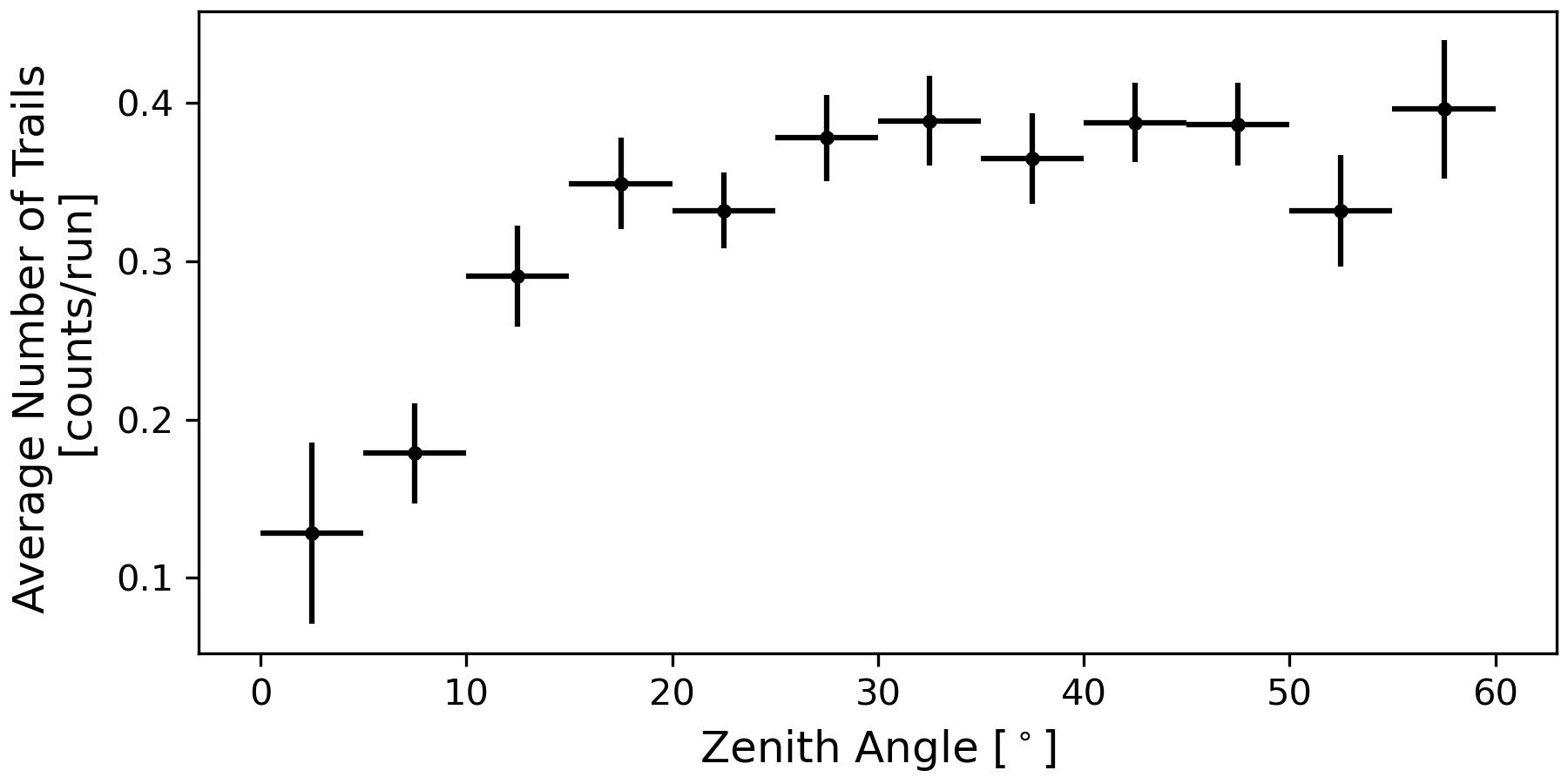}
\caption{The number of satellite trails we detect as a function of run start time and zenith angle.}
\label{fig:TINzen}
\end{figure}
Figure \ref{fig:monthly_avgerge} shows the number of trails we detect per month over time, excluding February 2022 as it biased our analysis (due to many observations of $\eta$ Carinae). We compare this to the number of satellites in orbit taken from the Union of Concern Scientists (UCS) satellites database \cite{UCS}. As this data ends in April 2022, we augment this dataset with later constellation launch data from McDowell \cite{mcdowell} as a lower limit for the total number in orbit. The number of trails we detect over time is mildly increasing (with a linear gradient fit inconsistent with zero), and mildly correlated (Pearson correlation coefficient of 0.4) with our combined satellite launch dataset. 

Figure \ref{fig:TINzen} shows the number of trails we detect as a function of run start time and zenith angle. We detect notably more trails at the beginning and end of the observing night, as this is when low-altitude satellites are most visible. High-zenith angle observations also contain more trail detections due to an increasing effective sampling volume \cite{Bassa}. Approximately $\sim0.2\%$ of dark observing time is contaminated by satellite trails in our dataset. Even if we were to be conservative and exclude all of this trail-contaminated data from H.E.S.S. analyses, it would not have a significant effect on the scientific performance of H.E.S.S..

\begin{figure}
    \centering
    \includegraphics[width = \hsize]{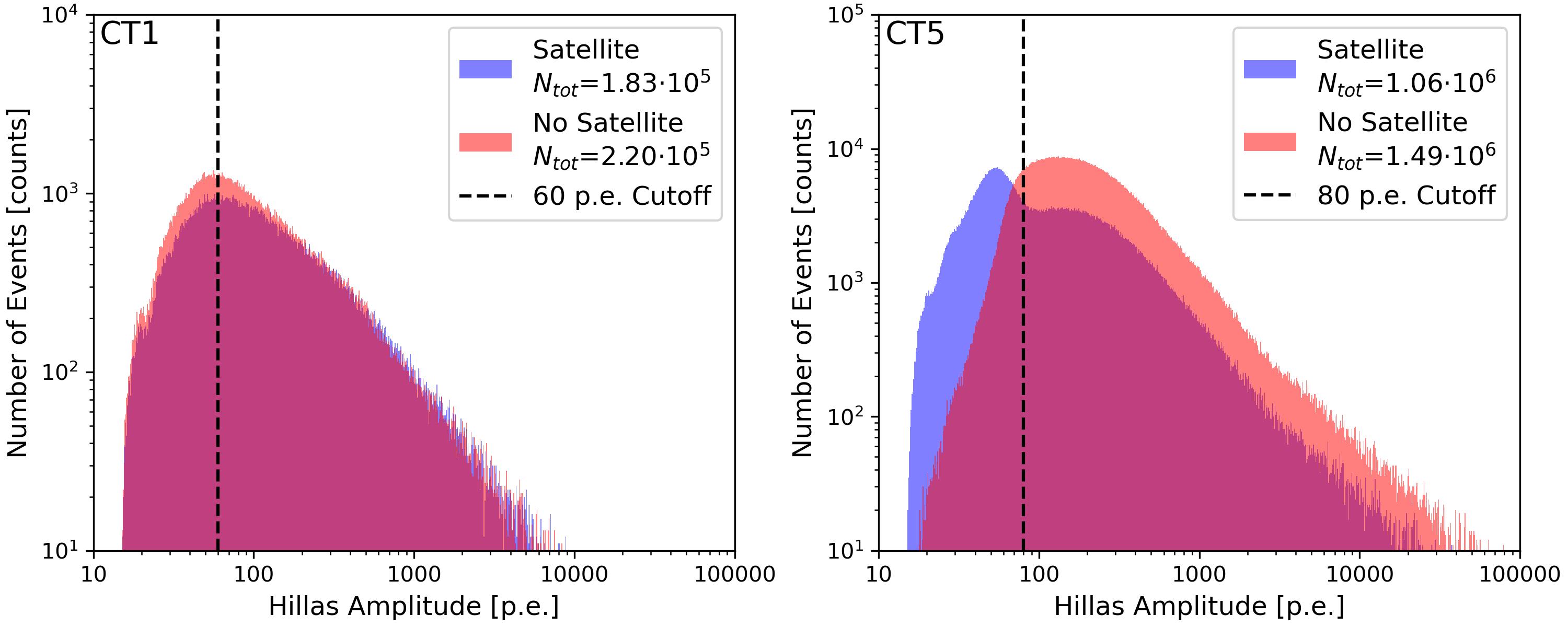}
    \caption{Hillas amplitude values for the brightest 99 trails we detect in CT1 and CT5, standard parameter cuts used in H.E.S.S. Heidelberg Analysis Pipeline analyses are shown for comparison. Looser cuts (reduced by $\mathrm{20\,p.e.}$ per-telescope) are occasionally used to lower the effective energy threshold of the instrument, such as during observations of gamma-ray busts \cite{221009A}.}
    \label{fig:amps}
\end{figure}

\begin{figure}
    \centering
    \includegraphics[width = \hsize]{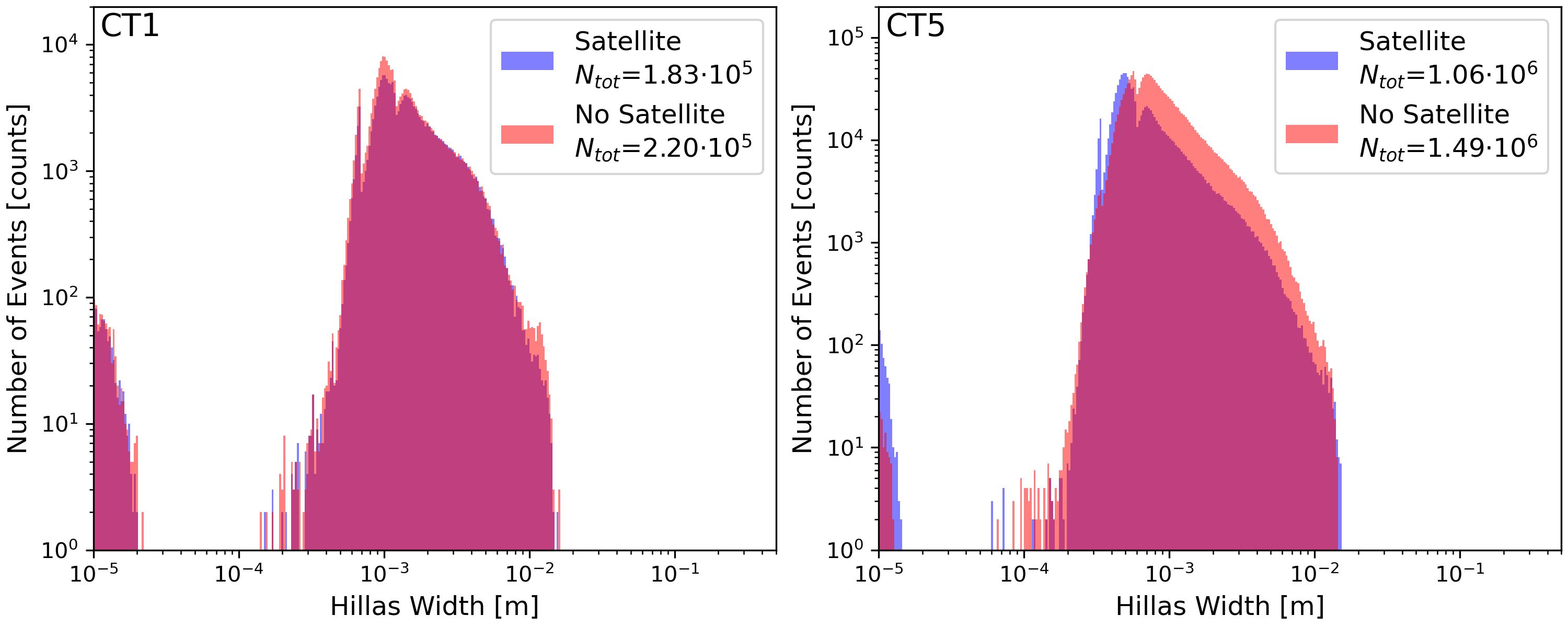}
    \caption{Hillas widths for the brightest 99 trails we detect in CT1 and CT5. Largely unaffected Hillas parameter distributions for the dimmer tracks can be found in \cite{Langetal}.}
    \label{fig:widths}
\end{figure}

In order to consider the effect of the trails upon event classification and reconstruction, we examine Hillas parameter values for events taken during equal periods of time during the trail and outside of the trail. This is as the satellite trails affect the trigger system of CT5; we observed that during the brightest trails the trigger rate in CT5 could drop suddenly (due to activation of safety mechanisms to prevent the camera from triggering too often) whilst the trigger rate in the other cameras increased. Examples of these distributions for the brightest $\sim5\%$ of trails observed with CT1 (as an example of a smaller H.E.S.S. telescope) and CT5 are seen in Figures \ref{fig:amps} and \ref{fig:widths}. The most notable effect we observed was a propensity for NSB-like false trigger events with a low Hillas amplitude in the CT5 camera, though such events would be removed by standard analysis cuts performed during H.E.S.S. analyses. But these events would likely not be removed and could potentially contaminate a scientific analysis if a new technique aiming to reduce the energy threshold of the instrument was used (such as deep-learning-based event classification \cite{shilon}). The effect on Hillas widths and other Hillas parameters (such as Hillas alpha) used for background rejection and directional reconstruction appears to be minimal in both CT1 and CT5. 

CTA will consist of three classes of IACT on two sites, one near Cerro Paranal in Chile and the other on the Spanish island of La Palma. Rough estimates of the number of trails that will be detected by CTA and the dark observing time contaminated can be made by (for example) extrapolating the linear trends we observe in Figure \ref{fig:monthly_avgerge} and scaling the predictions based on the Field of View of the telescope (relative to a value of $\mathrm{3.4^{\circ}}$ for FlashCam on CT5). These are shown in Table \ref{table:pred}. A non-trivial percentage of CTA observing time will be affected by the presence of trails, but our results suggest only a small fraction of this data would be sufficiently affected to justify its removal from scientific analysis. However, such predictions are subject to large uncertainties, particularly associated with instrument design. As such, other current generation facilities should perform similar studies to ours to verify our conclusions. That said, our results suggest that, for example, studies into modifying the CTA observation auto-scheduler to account for the presence of trails (such as have been performed for the Vera Rubin Observatory \cite{Huetal}) are unlikely to be required.

\begin{table}
\caption{Predictions for the number of trails per hour and fraction of observing time affected for CTA in 2030, based on assuming the linear trends we observe in H.E.S.S. data continue, and simple FoV scaling . These predictions are subject to significant uncertainties, and should be treated only as an approximate estimate.}
\label{table:pred}      
\centering          
\begin{tabular}{c c c c}     
\hline
Telescope Class & FoV Size [$^{\circ}$] & Average Trails Per Hour [counts] & Observing Time  Affected [\%]\\ \hline
Small & 9.0 & 15 & 5\\
Medium & 8.0 & 12 & 4\\   
Large & 4.3 & 3 & 2\\
\hline
\end{tabular}
\end{table}

\section{Conclusions}
We have developed a new technique to detect satellite trails with IACTs, and performed a search for trails in approximately three years of H.E.S.S. data. The number of trails we detect is rising over time, and is mildly correlated with the number of launches, suggesting the fraction of CTA data in which trails will appear is non-trivial. Our analysis has also demonstrated that the effect on IACT event classification and reconstruction from satellites is minimal. CTA could, however, be more affected if new analysis techniques that aim to reduce the energy threshold of the detector are applied. 

\section{Acknowledgements}
This work has been through review by the H.E.S.S. collaboration, who we thank for allowing us to use low level H.E.S.S. data in this work, and for useful discussions with collaboration members regarding this paper. This work is supported by the Deutsche Forschungsgemeinschaft (DFG, German Research Foundation) – Project Number 452934793.

%
%
%

\end{document}